\documentclass[letters,usenatbib]{mn2e}

\usepackage{graphicx}
\usepackage{multirow}
\usepackage{txfonts}

\usepackage{natbib}
\title[Implementation of TCAF Solution in XSPEC]
{Implementation of Two Component Advective Flow Solution in XSPEC}
\author[D. Debnath, S. K. Chakrabarti, $\&$ S. Mondal]
       {Dipak Debnath$^1$\thanks{E-mail: dipak@csp.res.in}, Sandip K. Chakrabarti$^{1,2}$, Santanu Mondal$^1$\\
$^1$ Indian Centre For Space Physics, 43 Chalantika, Garia Station Road, Kolkata, 700084, India\\
$^2$ S. N. Bose National Center for Basic Sciences, JD-Block, Salt Lake, Kolkata, 700098, India}

\date{Received: 2013 August 22; Accepted: 2014 February 3}

\begin{document}

\maketitle

\begin{abstract}
Spectral and Temporal properties of black hole candidates can be explained reasonably  
well using Chakrabarti-Titarchuk solution of two component advective flow (TCAF). This model requires 
two accretion rates, namely, the Keplerian disk accretion rate and the halo accretion rate, the latter  
being composed of a sub-Keplerian, low angular momentum flow which may or may not develop a shock.
In this solution, the relevant parameter is the relative importance of the halo (which creates the Compton 
cloud region) rate with respect to the Keplerian disk rate (soft photon source). 
Though this model has been used earlier to 
manually fit data of several black hole candidates quite satisfactorily, for the first time, 
we made it user friendly by implementing it into XSPEC software of GSFC/NASA. This enables any user 
to extract physical parameters of the accretion flows, such as two accretion rates, 
the shock location, the shock strength etc. for any black hole candidate. 
We provide some examples of fitting a few cases using this model. Most importantly, unlike any other 
model, we show that TCAF is capable of predicting timing properties from the spectral fits, since in TCAF, 
a shock is responsible for deciding spectral slopes as well as QPO frequencies.
\end{abstract}

\begin{keywords}
X-Rays: binaries, Stars:individual (H~7143-322, GX~339-4, GRO~J165540), Black Holes, Spectrum, Accretion disks, Shock waves, Radiation hydrodynamics
\end{keywords}

\section{Introduction}

Compact objects, such as black holes, neutron stars, etc. are identified by electromagnetic radiations 
emitted from the accreting matter. Understanding the spectral and timing properties of this radiation is 
essential for model builders and theorists alike. In a binary system, matter from the companion star 
accrets into the black hole through the Roche lobe, and/or through capturing its winds. This matter 
produces a disk like structure around the primary compact object. There are a large number of theoretical 
models in the literature which explain the physics of accretion around a black hole. Evidences of the 
standard disk proposed by \citet{SS73} and \citet{NT73} 
are present in most of the binary systems. However, the emitted spectrum of the radiation is multi-color 
in nature and contains both thermal and non-thermal components and a standard disk cannot explain the entire 
X-ray spectral features. Moreover, the inner region of the standard disk may be unstable due to the viscous 
and thermal effects \citep{Lightman74,Kobayashi03}. 

Simply put, one of the components of the spectrum is a multi-color blackbody radiation from the standard 
Keplerian disk and the other is a power-law component formed due to repeated Compton scatterings of the 
low energy (soft) photons of this blackbody by the hot electrons of the `Compton' cloud 
\citep{ST80,ST85}. 
There are many speculations regarding the nature of this Compton cloud ranging from a magnetic corona 
\citep{Galeev79}, to hot gas corona over the disk \citep{Haardt93,Zdziarski03}. 
Since the formation process of a static corona around an accretion disk is totally unknown, and since 
a low angular dynamic flow may naturally act as a 
corona, \citet[][hereafter CT95]{CT95} 
proposed that a disk having two distinct components, a Keplerian disk submerged inside a sub-Keplerian 
halo is enough to explain all the spectral properties very satisfactory. Observational evidences also 
started to support this so-called two-component 
advective flow (TCAF) \citep[e.g.,][]{Soria01,Smith02,Wu02,Cambier13}. 
While creating a self-consistent TCAF solution, the properties of a viscous transonic flow was made use of, 
in which a flow having viscosity above a critical value naturally forms a Keplerian disk and the region with 
a lower viscosity, due to centrifugal barrier, forms a shock wave, typically, at a few tens of Schwarzschild radii. 
The post-shock region (from the shock and the inner sonic point) basically evaporates the Keplerian component 
and together acts as a Compton cloud which produces a power-law component (hard photons) with exponential 
cut-off in the spectrum through thermal Comptonization. From the inner sonic point 
to the horizon of the black hole (bulk motion 
dominated advective flow or BDAF) the matter is advected rapidly to the black hole. The bulk motion in this 
region also up-scatters the soft photons and produces a second power-law component even when the temperature 
of the region is zero. If the centrifugal barrier is not strong enough, the shock may not form, but the flow 
still slows down. The spectral properties in this case are discussed in \citet[][hereafter C97]{C97}. 
The CENtrifugal pressure supported BOundary Layer or CENBOL, referred to the post-shock region or centrifugal 
force dominated region, which is also the base of the outflows where the pre-Jet is launched, plays the most 
important role in the black hole physics. As usual, this CENBOL, pre-Jet and BDAF intercept soft photons from 
the Keplerian disk and reprocess them to high energies via inverse Compton scattering. In this {\it letter}, 
we will implement the TCAF solution to study spectral properties of black hole candidates (BHCs) using widely 
used user-friendly spectral analysis software package, developed by GSFC/NASA, called XSPEC. For the sake of 
concreteness, we focus only the cases where only the CENBOL is present. We ignore the effects of BDAF and 
pre-Jet. In our next version of analysis, these components and spin of the black hole would be included.

The Galactic transient black hole candidates are very interesting objects to study in X-rays because 
these sources generally show rapid evolutions in their temporal and spectral properties during their outburst 
phases, which are strongly correlated to each other \citep[see for a review,][]{RM06}. 
In general, four basic states - {\it hard, hard-intermediate, soft-intermediate}, and {\it soft} states are 
observed during an outburst of the BHCs \citep[see,][and references therein]{Nandi12}. The evolutions of 
these spectral states are observed, which indeed form a hysteresis loop during the outburst with hard states 
are found to be at the beginning and end time of the outbursts, whereas soft and intermediate spectral states 
are observed in between. The evolution of spectral states are strongly dependent on the variation of the 
accretion rates. According to the TCAF solution, accretion flow rates may be controlled by a physical parameter, 
such as the magnetic viscosity, perhaps owing to the enhanced magnetic activity of the companion 
\citep{Wu02,Nandi12,DD13}.       
During the rising phase of the outburst, viscosity may cause an increase in the accretion rate of the Keplerian 
matter. As the viscosity is reduced, the Keplerian rate is reduced and declining phase starts. The Keplerian disk 
itself recedes away leaving behind only the low-angular sub-Keplerian flow causing a hard state. 
Thus, a rigorous fit with TCAF model is expected to throw light on how the accretion rates and the flow 
geometry evolve with time. 

In general, low and intermediate frequency quasi-periodic oscillations (LFQPOs) are observed in hard and 
intermediate (hard-intermediate and soft-intermediate) spectral states of transient black hole candidates. 
These QPOs are reported extensively in literature, although still there are debates on the origin of these QPOs.
However, according to the shock oscillation model (SOM) by Chakrabarti and his collaborators, LFQPOs are originated 
due to the oscillation of the post-shock region (\citealt{MSC96}, hereafter MSC96; \citealt{CAM04}, 
hereafter CAM04; \citealt{GGC14}, hereafter GGC14)
when the resonance occurs between the infall time scale and the cooling time scale in CENBOL. During oscillation,
the shape of the Compton cloud and the degree of interception change periodically. Since from our 
spectral fit, we can directly extract values of physical parameters related to this shock wave, we can also
predict what should be the frequency of the observed low frequency QPO (if present; see \S 4.1 for details).

The {\it paper} is organized in the following way: in the next Section, we briefly describe properties of 
the TCAF model. In \S 3, we discuss the method of the implementation of the TCAF model in XSPEC for spectral 
fittings. In \S 4, TCAF model fitted results obtained from the spectral fit of three different BHCs.
Finally, in \S 5, we make concluding remarks and our future work plans.

\section{A brief description of TCAF model}

The TCAF model (CT95, C97) has been described in detail in the literature and has been proven to be 
a stable configuration by extensive numerical simulations \citep{GC13}. 
The model requires two accretion rates: one is the rate of the Keplerian component and the other is the rate 
of the low-angular momentum, sub-Keplerian halo, in which Keplerian disk is immersed. Two other essential 
parameters are the shock location and the compression ratio of the flow at the shock respectively. 
These two parameters provide the height of the shock, calculated using the pressure balance condition \citep{C89}. 
The density and temperature distribution of the flow and especially in the 
post-shock region are calculated using two temperature equations and continuity equations 
as discussed in CT95. 
The CT95 code also computes the optical depth, average electron temperature of the CENBOL, the spectral 
index etc. self-consistently by adding the relevant cooling and heating processes such as arising 
due to bremsstrahlung, Comptonization, inverse bremsstrahlung and inverse Comptonization.
Synchrotron cooling process was not included in this version. CT95 considered only strong shock
case. In order to take care of the weaker shocks also, we generalized the expression for the 
shock height ($H_{shk}$) and shock temperature ($T_{shk}$) in the following way:
$$
H_{shk}=\left[\frac{\gamma (R-1) {X_{s}}^{2}}{R^{2}}\right]^{\frac{1}{2}}
\eqno{(1)}
$$
and the shock temperature is given by,
$$
T_{shk}=\frac{m_p (R-1)c^{2}}{2R^{2}k_{B}(X_{s}-1)}
\eqno{(2)}
$$
where, $m_p$, $R$, $k_{B}$ $X_{s}$ and $\gamma$ are the mass of the proton, compression ratio, Boltzmann 
constant, shock location and adiabatic constant of the flow respectively. We also incorporate the spectral 
hardening correction \citep[see,][hereafter DMC14]{DD14a} depending on the accretion flow rate as 
in \citet{ST95}. 
\citet{Paczynski80} 
pseudo-Newtonian potential $\Phi_{PN}=-\frac{1}{2(r-1)}$ has been used to describe the geometry around 
the black hole. 

\section{Procedure of Implementation of TCAF into XSPEC}

To fit a spectrum with the TCAF model using HEASARC's spectral analysis software package XSPEC, 
which already has a number of inbuilt theoretical models, we need to first generate a model {\it fits} 
file by varying five different input parameters: Keplerian rate (disk rate $\dot{m_d}$), sub-Keplerian rate 
(halo rate $\dot{m_h}$), mass of the black hole $M_{BH}$, location of the shock $X_s$, and the compression 
ratio $R$ and use it as a system model. 
In order to fit the spectra in XSPEC, we generated  a additive table model {\it fits} file named ({\it TCAF.fits}). 
We first incorporated changes as regards to shock strength as described above in the CT95 model code 
(for details see, DMC14) 
and generated $\sim 4\times10^5$ model spectra by solving the theoretical radiative-hydro code of CT95.
For each spectrum, we provide five input parameters by varying five input parameters ($\dot{m_d}$, $\dot{m_h}$, 
$M_{BH}$, $X_s$, and $R$) in the following ranges : i) $0.1 - 12.1$ $\dot{M}$$_{Edd}$, 
ii) $0.01 - 12.01$ $\dot{M}$$_{Edd}$, iii) $5 -15$ Solar mass ($M_\odot$), iv) $6 - 456$ $r_g$, and 
v) $1 - 4$, respectively. Here, $\dot{M}$$_{Edd}$ is the Eddington rate.
These model spectra are used as input files to a program 
written in FORTRAN, to generate the model {\it fits} file. 

At present, we have fitted the spectra after keeping model {\it fits} file as a local additive table model. 
At the time of spectral fitting using the TCAF, one needs to supply six model initial guess parameters: 
i) Keplerian rate ($\dot{m_d}$ in units of $\dot{M}$$_{Edd}$), ii) sub-Keplerian rate ($\dot{m_h}$ in units of 
$\dot{M}$$_{Edd}$), iii) black hole mass ($M_{BH}$) in units of $M_\odot$, iv) location of the shock ($X_s$ in 
units of Schwarzschild radius $r_g=2GM_{BH}/c^2$), v) compression ratio ($R=\rho_+ / \rho_-$, where $\rho_+$ and 
$\rho_-$ are densities of the post- and pre- shock matters) of the shock, and vi) the model normalization value 
($norm$), which is equivalent to $\frac{1}{4\pi D^2} cos(i)$, where $D$ is the source distance in $10$~kpc unit 
and $i$ is the disk inclination angle. In the near future, the fits file will be made public, for the use of the 
scientific community. It would be available now upon request.  

\section {Results: Sample Spectra Fitted with TCAF Model}

We now show the results of fitting of three $2.5-25$~keV background subtracted RXTE/PCA spectra of
three different black hole candidates, namely, H~1743-322, GX~339-4, GRO~J1655-40. These observations are 
taken from the initial phase of the outbursts, where QPOs are observed. We carry out data analysis using 
the FTOOLS software package HeaSoft version HEADAS 6.12 and XSPEC version 12.7. For the generation of source 
and background `.pha' files and spectral fittings using TCAF model we use the same method as mentioned in 
DMC14. 

It is to be noted that the TCAF model in its present form (i.e., without incorporating pre-Jet and BDAF) 
is able to fit hard and intermediate state spectra with acceptable values of reduced $\chi^2$ ($\leq 2$).  
In the soft states, the shock does not form, and the inclusion of BDAF is required (see, DMC14) 
for an acceptable fit. As a result, the  number of parameters is required, will be reduced to three. 
This will be carried out in the  near future.

\begin{figure}
\vskip -0.3 cm
\centering{
\includegraphics[scale=0.6,angle=270,width=4.0truecm]{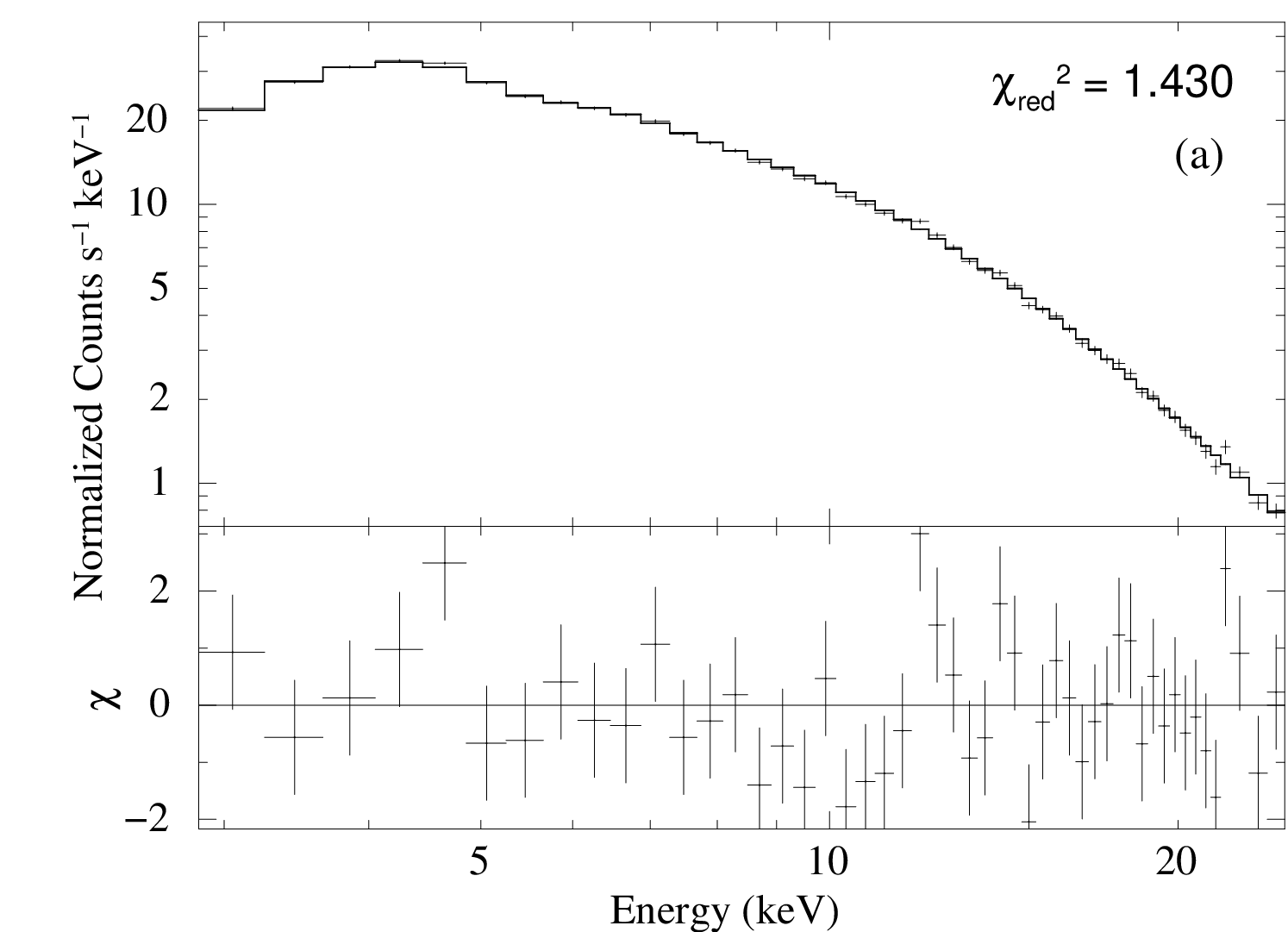}
\includegraphics[scale=0.6,angle=270,width=4.0truecm]{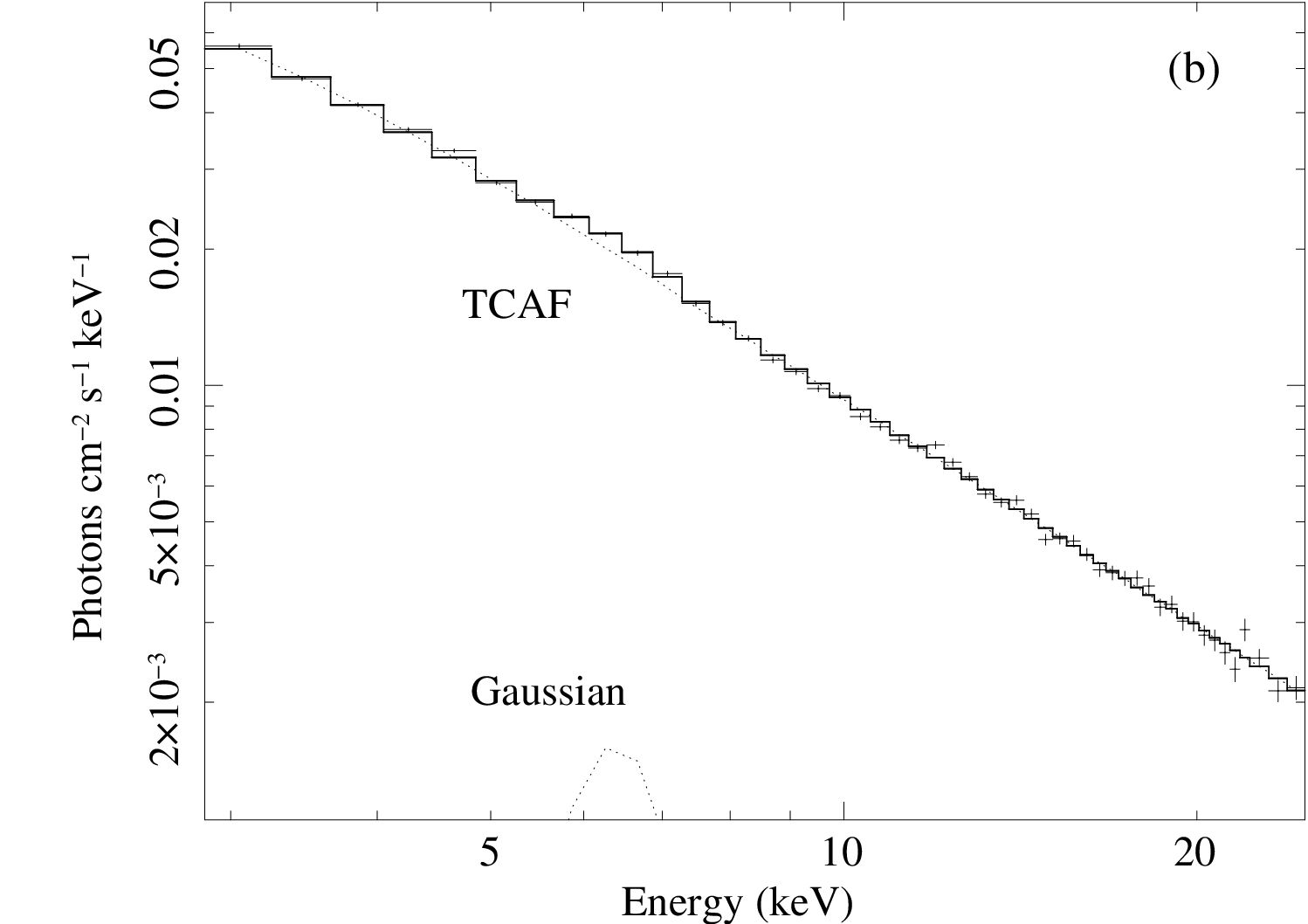}}
\caption{(a) TCAF model fitted $2.5-25$~keV PCA spectrum of H~1743-322 
(Observation ID = 95360-14-02-01; MJD = 55419) with variation of $\Delta \chi$ is shown in 
the left panel. The value of the model fitted reduced $\chi^2$ is written down. 
(b) The unfolded model components of the spectral fit is shown in the right panel.
}
\label{kn : fig1}
\end{figure}

In Fig. 1, $2.5-25$~keV background subtracted PCA spectrum from Galactic transient BHC H~1743-322
of observation ID = 95360-14-02-01 (MJD = 55419.1070) from its 2010 outburst is shown. Fixed values 
of 1\% systematic error, the hydrogen column density ($N_H$) of $1.6 \times 10^{22}$ \citep{DD13} 
for absorption model {\it wabs}, and $M_{BH}$ of $11.4\pm1.9$ \citep{DD14b} are used to fit the spectrum. 
To achieve the best fit, a single Gaussian Iron line $6.39\pm0.19$~keV is also used. With these, the 
value of the reduced $\chi^2 = 1.430$ is achieved. For details about the model fitted parameters, see Table 1.

\begin {figure}
\vskip -0.3 cm
\centering{
\includegraphics[scale=0.6,angle=270,width=4.0truecm]{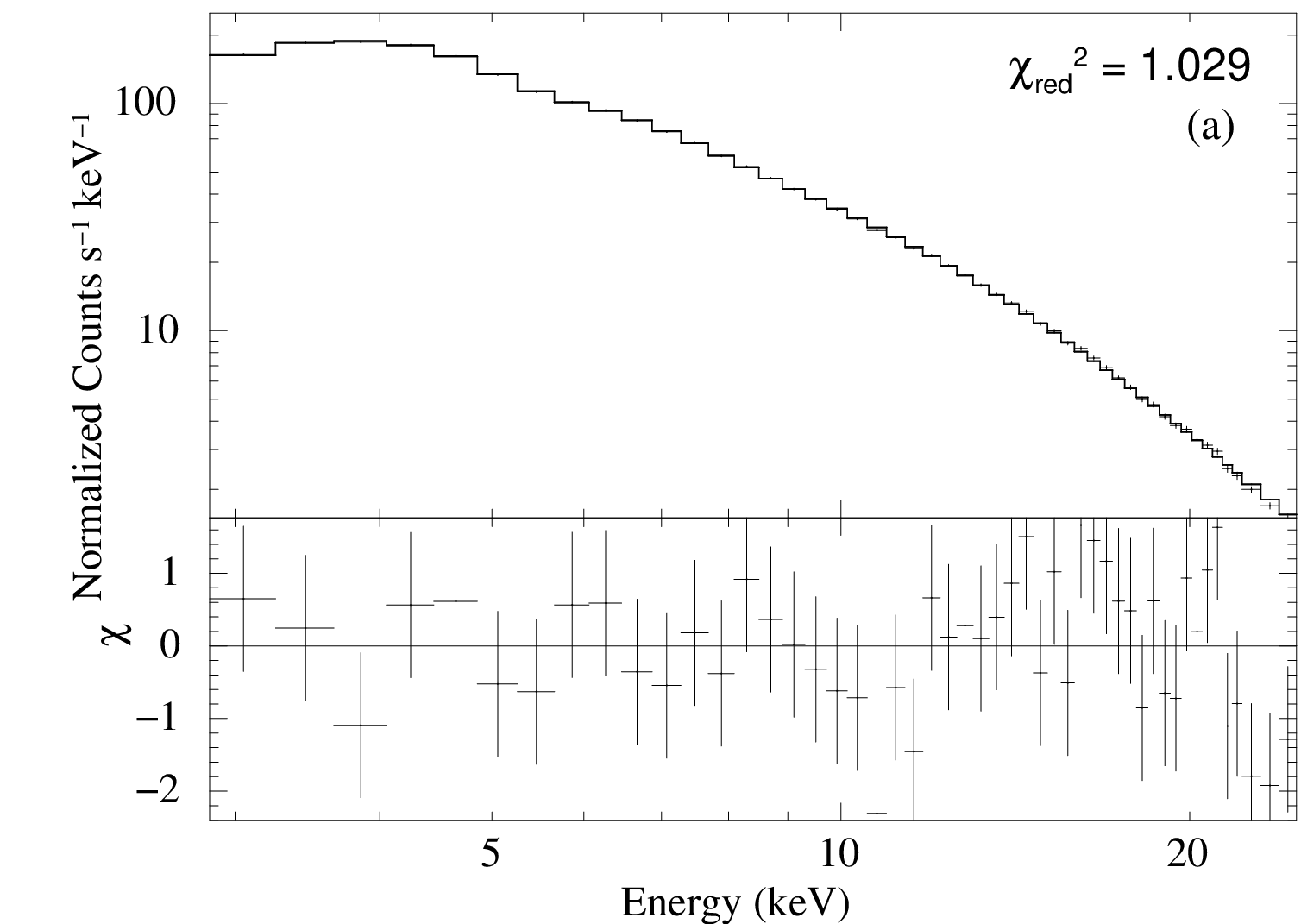}
\includegraphics[scale=0.6,angle=270,width=4.0truecm]{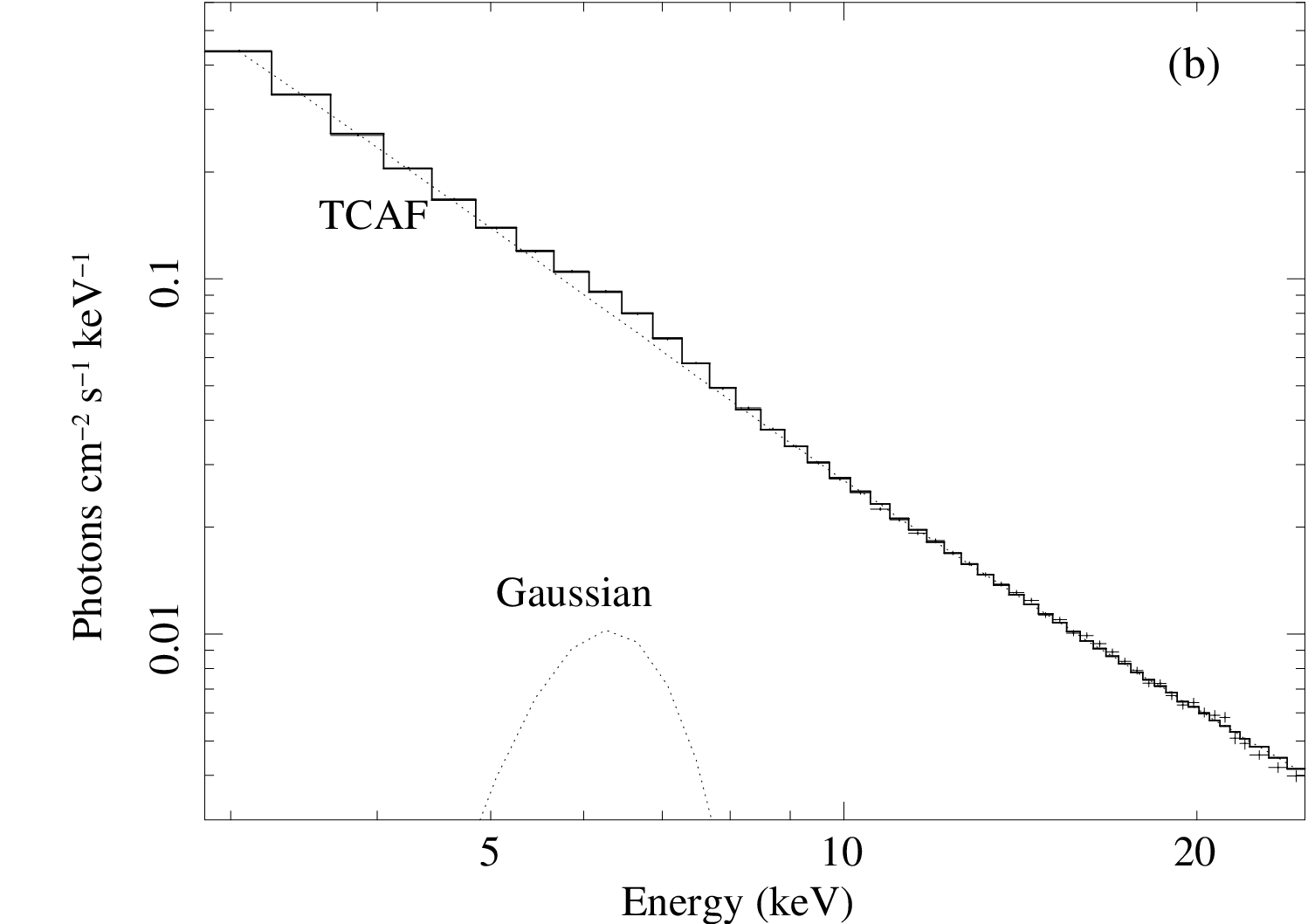}}
\caption{Same as of Fig. 1(a-b), except the spectrum of 
GX~339-4 (Observation ID = 95409-01-14-04; MJD = 55300) is used.
}
\label{kn : fig2}
\end {figure}

In Fig. 2, RXTE/PCA spectrum of Galactic outbursting BHC GX~339-4 fitted with combination of TCAF and 
single Gaussian line ($6.31\pm0.17$~keV) is shown. This observation (ID = 95409-01-14-04; MJD = 55300.3421) 
is selected from the rising phase of 2010-11 outburst of GX~339-4. For the spectral 
fitting, fixed values of $M_{BH}$ = $5.8\pm0.5$ \citep{Hynes03}, $N_H$ = $5 \times 10^{21}$ \citep{DD10} 
for absorption model {\it wabs} and 1\% systematic error are used. In this case, a value of the 
reduced $\chi^2 = 1.029$ is achieved.

\begin {figure}
\vskip -0.3 cm
\centering{
\includegraphics[scale=0.6,angle=270,width=4.0truecm]{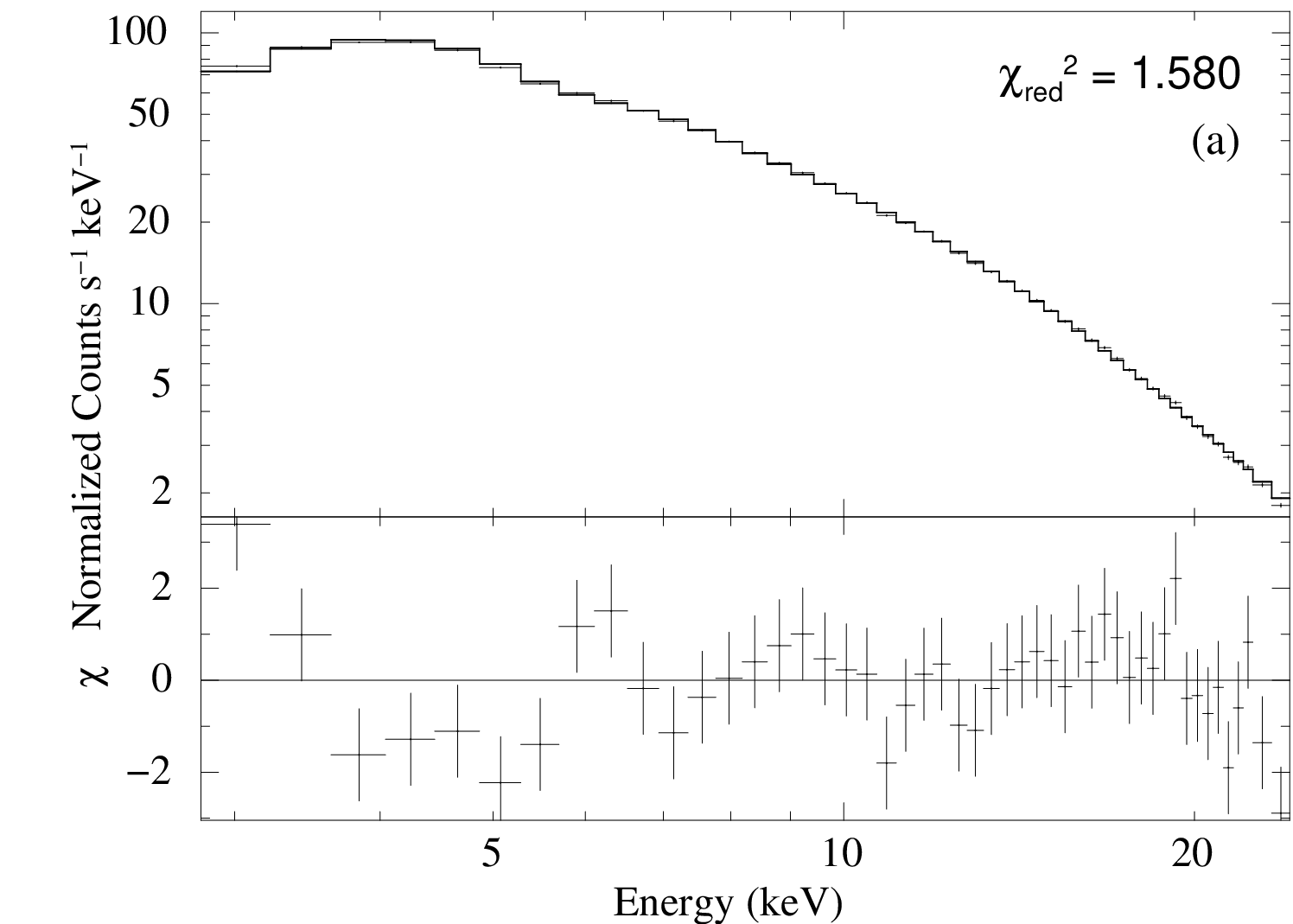}
\includegraphics[scale=0.6,angle=270,width=4.0truecm]{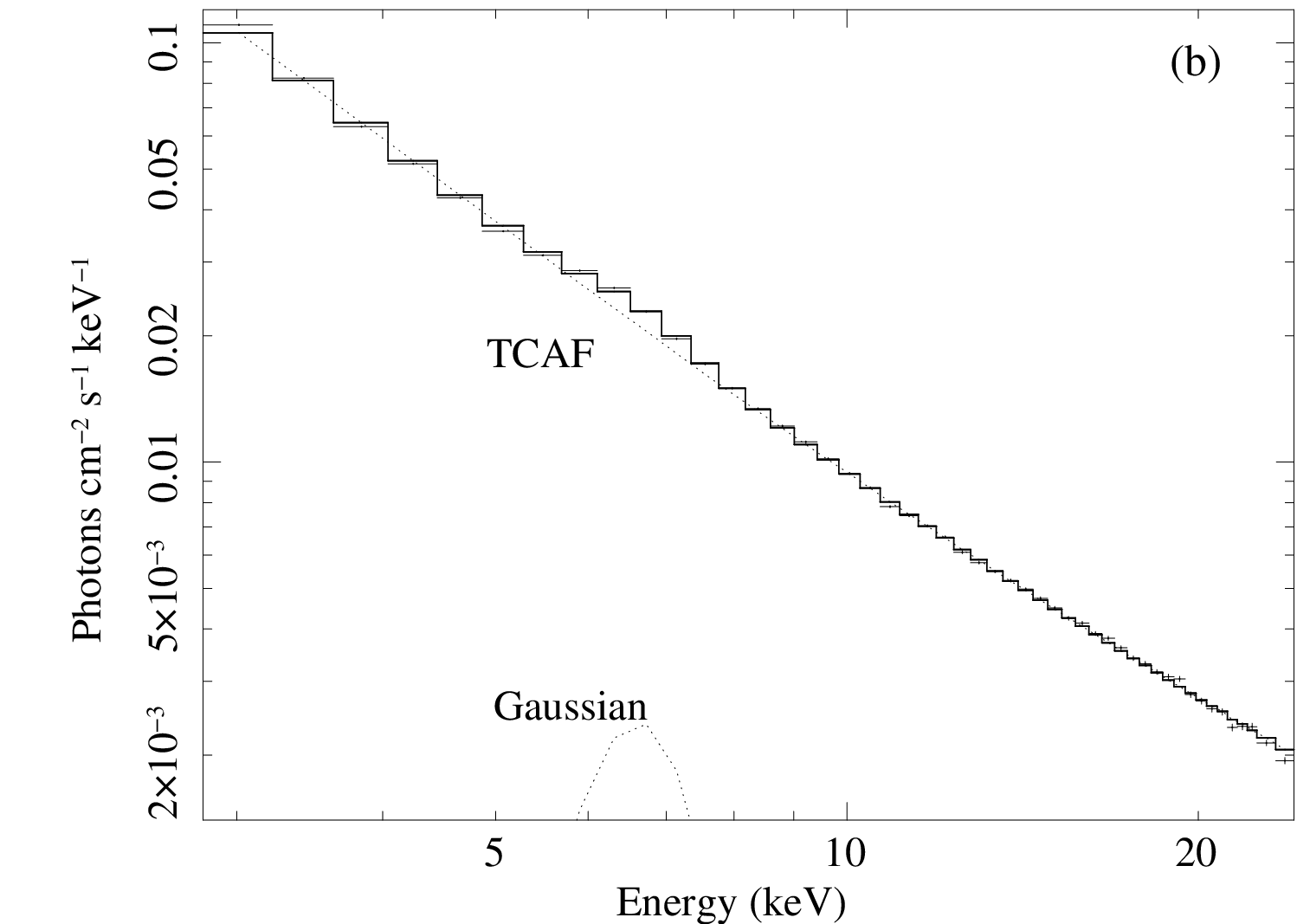}}
\caption{Same as of Fig. 1(a-b), except the spectrum of GRO~J1655-40 
(Observation ID = 90704-04-01-00; MJD = 53439) is used.
}
\label{kn : fig3}
\end {figure}

In Fig. 3, TCAF model fitted RXTE/PCA spectrum from the 2005 outburst of the Galactic outbursting BHC 
GRO~J1655-40 is shown. The spectrum is fitted with combination of two additive model components, namely, 
TCAF and a Gaussian line ($6.62\pm0.15$~keV). This observation of ID = 90704-04-01-00 (MJD = 53439.7603) 
is selected from the rising phase of 2005 outburst of the source. For the spectral fitting, fixed values 
of $M_{BH}$ equals to $7.02\pm0.22$ \citep{Orosz97}, $N_H$ equals to $7.5 \times 10^{21}$ \citep{DD08} 
for absorption model {\it wabs} and 1\% systematic error are used. In this case, a value of the reduced 
$\chi^2 = 1.580$ is achieved. 

\begin{table}
\addtolength{\tabcolsep}{-4.50pt}
\vskip -0.4 cm
\small
\centering
\caption{\label{table1} TCAF Model Fitted Spectral Result}
\vskip -0.2cm
\begin{tabular}{|l|cccccccc|}
\hline
Source & Obs. Id & $\dot{m_d}$ & $\dot{m_h}$ & $X_s$ & R &$ \chi^2$/DOF & $\nu_{QPO}^*$ & $\nu_{QPO}^*$ \\
       &         & ($\dot{M}$$_{Edd}$) & ($\dot{M}$$_{Edd}$) &($r_g$)&   &              & (Obs.) & (Predic.) \\
\hline
H 1743-322  &X-02-01& 0.516    & 0.189    & 320.0   & 1.250    & 60.1/42 & 1.045    & 1.228 \\
            &       &$\pm0.013$&$\pm0.081$&$\pm20.04$&$\pm0.012$&         &$\pm0.007$&$\pm0.293$ \\
GX 339-4    &Y-14-04& 6.883    & 6.087    & 147.9   & 4.000    & 43.2/42 & 2.374    & 2.356 \\
            &       &$\pm0.003$&$\pm0.349$&$\pm1.13 $&$\pm0.075$&         &$\pm0.006$&$\pm0.265$ \\
GRO J1655-40&Z-01-00& 6.987    & 1.733    & 153.8   & 3.449    & 64.78/41 & 2.313    & 2.172 \\
            &       &$\pm0.273$&$\pm0.232$&$\pm13.36$&$\pm0.433$&         &$\pm0.010$&$\pm0.529$ \\
\hline
\end{tabular}
\leftline {Here, X=95360-14, Y=95409-01, and Z=90704-04. DOF means degrees of freedom.}
\leftline {$^*$ Only frequency of the primary dominating QPOs (in Hz) are mentioned.}
\end{table}

\subsection{Prediction of QPO frequencies from the spectral fits using TCAF model}

Unlike any other model fits, TCAF model predicts timing properties from spectral fits. This is because the 
same shock which defines the CENBOL boundary, i.e., the size of the Compton cloud, 
also causes low frequency QPOs as it oscillates. 
The presence of a shock wave does not always mean for the existence of QPOs. 
The shock oscillation takes place provided the cooling and the infall times scales are of same order 
(MSC96, CAM04, GGC14) 
or when the Rankine-Hugoniot relation is not satisfied even with two sonic 
points in the transonic sub-Keplerian flow \citep{RCM97}. 

The frequency of oscillation is inversely proportional to the infall time ($t_{infall}$) in post-shock region 
(see, Eqn. 3 below) when the cooling time scale is also similar. One can determine the QPO frequency 
($\nu_{QPO}$) if the location of shock ($X_s$ in $r_g$) and the compression 
ratio ($R$) are known (see, Eqn. 4 below). 
In the presence of a shock, the infall time in the post-shock region can be expressed as, 
$$
t_{infall} \sim  R~X_s(X_s-1)^{1/2}.
\eqno{(3)}
$$
The frequency of the observed QPOs becomes, 
$$
\nu_{QPO} \sim t_{infall}^{-1} = C / [R~X_s(X_s-1)^{1/2}] ,
\eqno{(4)}
$$
where, $C$ is a constant = $M_{BH} \times 10^{-5}$. This shows that the derived $R$ and $X_s$ from 
the spectral fit leads to an estimate of the QPO frequency. 

In all of the three spectra fitted in this paper, QPOs are observed. From the spectral fit, we have therefore 
estimated the frequency of the QPOs, which roughly match with the observed values (see, Table 1). This is 
unique in the context of model fits.

\section {Concluding Remarks and Future Plan}

In this {\it letter}, we show how to implement the TCAF model in XSPEC as a local additive table model. In Figs. 1-3, 
we show the model fitted spectra, one each for the black hole candidates H~1743-322, GX~339-4, GRO~J1655-40 
respectively. We show that the TCAF model is quite capable to fit the black hole spectra. Moreover, fitting with 
TCAF model appears to be better than other conventional black body and power-law models because it can directly 
provide accretion rates from the spectral fit. The iterative procedure of CT95 also ensures that a 
no X-ray component reflected from the disk is required to be added. 
Not only that, unlike other models, TCAF has a predictive capability of the timing 
properties from the spectral fitted parameters. This is possible, because the same shock which decides the 
size of the Compton cloud parameters such as the optical depth and its average 
electron temperature (and thus the spectral index), also decides the QPO frequency. 
Detailed spectral study using TCAF model for the 2010-11 outburst of GX~339-4 (DMC14), 
and 2010 outburst of H~1743-322 \citep{Mondal14} will be published elsewhere. Detailed study on the evolution 
of QPO frequencies during the rising and the declining phases of the outburst will also be published elsewhere 
\citep{DD14c}. 

The present version of TCAF which is implemented here does not include the subsonic pre-Jet which is originated 
from the CENBOL. Similarly, it does not include the innermost bulk motion dominated region of the advective flow 
(BDAF) whose effect would be to produce a power-law component even if the Compton cloud is cooled down. These will 
enable us to fit not only the very soft states, but also those states with broken power-law as well as the jet 
dominated flows. We have verified that our present model fits hard and intermediate states very satisfactorily. 
The work to extend the validity of TCAF is in progress and would be reported elsewhere.

\noindent {\bf Acknowledgments :} 
We are thankful to NASA/GSFC scientists and XTEhelp team members (specially to Dr. Keith A. Arnaud) 
for their kind help to write FORTRAN programs to generate model {\it fits} file by using theoretical 
(TCAF) model spectra. 
We also acknowledge Mr. Sudip Garain of S. N. Bose National Centre for Basic Sciences for helps
in producing the large fits file. 
SM acknowledges financial supports from  CSIR (NET) scholarship.


{}


\begin{thebibliography}{}

\bibitem[\protect\citeauthoryear{Cambier \& Smith}{2013}]{Cambier13} Cambier, H.J., \&  Smith, D.M., 2013, ApJ, 767, 46
\bibitem[\protect\citeauthoryear{Chakrabarti}{1989}]{C89} Chakrabarti, S.K., 1989, MNRAS, 240, 7
\bibitem[\protect\citeauthoryear{Chakrabarti \& Titarchuk}{1995}]{CT95} Chakrabarti, S.K., \& Titarchuk, L.G., 1995, ApJ, 455, 623 (CT95)
\bibitem[\protect\citeauthoryear{Chakrabarti}{1997}]{C97} Chakrabarti, S.K., 1997, ApJ, 484, 313 (C97)
\bibitem[\protect\citeauthoryear{Chakrabarti, Acharyya, \& Molteni}{2004}]{CAM04} Chakrabarti, S.K., Acharyya, K., \& Molteni, D., 2004, A\&A, 421, 1 (CAM04)
\bibitem[\protect\citeauthoryear{Debnath et al.}{2008}]{DD08} Debnath, D., Chakrabarti, S.K., Nandi, A., \& Mandal, S., 2008, BASI, 36, 151
\bibitem[\protect\citeauthoryear{Debnath et al.}{2010}]{DD10} Debnath, D., Chakrabarti, S.K., \& Nandi, A., 2010, A\&A, 520, 98
\bibitem[\protect\citeauthoryear{Debnath et al.}{2013}]{DD13} Debnath, D., Chakrabarti, S.K., \& Nandi, A., 2013, AdSpR, 52, 2143
\bibitem[\protect\citeauthoryear{Debnath, Mondal \& Chakrabarti}{2014a}]{DD14a} Debnath, D., Mondal, S., \& Chakrabarti, S.K., 2014a, ApJ (submitted) (DMC14) (arXiv:astro-ph/1306.3745)
\bibitem[\protect\citeauthoryear{Debnath et al.}{2014b}]{DD14b} Debnath, D., et al., 2014b, MNRAS (in preparation)
\bibitem[\protect\citeauthoryear{Debnath et al.}{2014c}]{DD14c} Debnath, D., et al., 2014c, MNRAS (in preparation)
\bibitem[\protect\citeauthoryear{Ebisawa,Titarchuk \& Chakrabarti}{1996}]{Ebisawa96} Ebisawa, K., Titarchuk, L.G., \& Chakrabarti, S.K., 1996, PASJ, 48, 59
\bibitem[\protect\citeauthoryear{Galeev, Rosner \& Vaiana}{1979}]{Galeev79} Galeev, A.A., Rosner, R., \& Vaiana, G.S., 1979, ApJ, 229, 318
\bibitem[\protect\citeauthoryear{Garain, Ghosh \& Chakrabarti}{2014}]{GGC14} Garain, S., Ghosh, H., \& Chakrabarti, S.K., 2014, MNRAS, 437, 1329 (GGC14)
\bibitem[\protect\citeauthoryear{Giri \& Chakrabarti}{2013}]{GC13} Giri, K., \& Chakrabarti, S.K., 2013, MNRAS, 430, 2836
\bibitem[\protect\citeauthoryear{Haardt \& Maraschi}{1993}]{Haardt93} Haardt, F., \& Maraschi, L., 1993, ApJ, 413, 507
\bibitem[\protect\citeauthoryear{Hynes et al.,}{2003}]{Hynes03} Hynes, R. I., Steeghs, D., \& Casares, J., et al., 2003, ApJ, 583, L95
\bibitem[\protect\citeauthoryear{Kobayashi et al.}{2003}]{Kobayashi03} Kobayashi, Y., Kubota, A., Nakazawa, K., Takahashi, T., \& Makishima, K., 2003, PASJ, 55, 273
\bibitem[\protect\citeauthoryear{Lightman \& Eardley}{1974}]{Lightman74} Lightman, A. P., \& Eardley, D. M., 1974, ApJ, 187, 1
\bibitem[\protect\citeauthoryear{McClintock \& Remillard}{2003}]{MR03} McClintock, J.E., \& Remillard, R.A., 2003, arXiv, 036213 (astro-ph/0306213)
\bibitem[\protect\citeauthoryear{Molteni, Sponholz \& Chakrabarti}{1996}]{MSC96} Molteni, D., Sponholz, H., \& Chakrabarti, S.K., 1996, ApJ, 457, 805 (MSC96)
\bibitem[\protect\citeauthoryear{Mondal, Debnath \& Chakrabarti}{2014}]{Mondal14} Mondal, S., Debnath, D., \& Chakrabarti, S.K., 2014, ApJ (submitted) (arXiv:astro-ph/1401.4239)
\bibitem[\protect\citeauthoryear{Nandi et al.}{2012}]{Nandi12} Nandi, A., Debnath, D., Mandal, S., \& Chakrabarti, S.K., 2012, A\&A, 542, 56
\bibitem[\protect\citeauthoryear{Novikov \& Thorne}{1973}]{NT73} Novikov, I., \& Thorne, K.S., 1973, in Black Holes, Ed. C. DeWitt \& B.S. DeWitt (New York: Gordon \& Breach), 343
\bibitem[\protect\citeauthoryear{Orosz \& Bailen}{1997}]{Orosz97} Orosz, J.A., \& Bailen, C.D., 1997, ApJ 477, 876
\bibitem[\protect\citeauthoryear{Paczy\'nski \& Witta}{1980}]{Paczynski80} Paczy\'nski, B., \& Witta, P.J., 1980, A\&A, 88, 23
\bibitem[\protect\citeauthoryear{Remillard \& McClintock}{2006}]{RM06} Remillard, R.A., \& McClintock, J.E., 2006, ARA\&A, 44, 49
\bibitem[\protect\citeauthoryear{Ryu, Chakrabarti \& Molteni}{1997}]{RCM97} Ryu, D., Chakrabarti, S.K., \& Molteni, D., 1997, ApJ, 474, 378
\bibitem[\protect\citeauthoryear{Shakura \& Sunyaev}{1973}]{SS73} Shakura, N.I., \& Sunyaev, R.A., 1973, A\&A, 24, 337 (SS73)
\bibitem[\protect\citeauthoryear{Shimura \& Takahara}{1995}]{ST95} Shimura, T., \& Takahara, F., 1995, ApJ, 445, 780
\bibitem[\protect\citeauthoryear{Smith, Heindl \& Swank}{2002}]{Smith02} Smith, D., Heindl, W.A., \& Swank, J.H., 2002, ApJ, 569, 362
\bibitem[\protect\citeauthoryear{Soria et al.}{2001}]{Soria01} Soria, R., Wu, K., \& Hannikainen, D., et al., 2001, Proc. of workshop on ``X-Ray emission from Accretion onto Black Holes", 65
\bibitem[\protect\citeauthoryear{Sunyaev \& Titarchuk}{1980}]{ST80} Sunyaev, R.A., \& Titarchuk, L.G., 1980, ApJ, 86, 121
\bibitem[\protect\citeauthoryear{Sunyaev \& Titarchuk}{1985}]{ST85} Sunyaev, R.A., \& Titarchuk, L.G., 1985, A\&A, 143, 374
\bibitem[\protect\citeauthoryear{Wu et al.}{2002}]{Wu02} Wu, K. et al., 2002, ApJ, 565, 1161
\bibitem[\protect\citeauthoryear{Zdziarski et al.}{2003}]{Zdziarski03} Zdziarski, A.A., Lubinski, P., \& Gilfanov, M., et al., 2003, MNRAS, 342, 355




\end{thebibliography}
\end{document}